Far-Infrared Spectroscopy of Cationic Polycyclic Aromatic Hydrocarbons:

Zero Kinetic Energy Photoelectron Spectroscopy of Pentacene Vaporized from Laser Desorption

Suggested running head: Far-IR spectroscopy of pentacene cation

Jie Zhang, Fangyuan Han, Linsen Pei, Wei Kong*

Department of Chemistry, Oregon state University, Corvallis, OR 97331, USA

**and**

Aigen Li

Department of Physics and Astronomy, University of Missouri, Columbia, MO 65211, USA

*Corresponding author: Wei.Kong@oregonstate.edu

ABSTRACT

The distinctive set of infrared (IR) emission bands at 3.3, 6.2, 7.7, 8.6, and 11.3 μm are ubiquitously seen in a wide variety of astrophysical environments. They are generally attributed to polycyclic aromatic hydrocarbon (PAH) molecules. However, not a single PAH species has yet been identified in space as the mid-IR vibrational bands are mostly representative of functional groups and thus do not allow one to fingerprint individual PAH molecules. In contrast, the far-IR (FIR) bands are sensitive to the skeletal characteristics of a molecule, hence they are important for chemical identification of unknown species. With an aim to offer laboratory astrophysical data for the Herschel Space Observatory, Stratospheric Observatory for Infrared Astronomy, and similar future space missions, in this work we report neutral and cation FIR spectroscopy of pentacene ($C_{22}H_{14}$), a five-ring PAH molecule. We report three IR active modes of cationic pentacene at 53.3, 84.8 and 266 μm that may be detectable by space missions such as the SAFARI instrument on board SPICA.

In the experiment, pentacene is vaporized from a laser desorption source and cooled by a supersonic argon beam. We have obtained results from two-color resonantly enhanced multiphoton ionization (REMPI) and two-color zero kinetic energy photoelectron spectroscopy (ZEKE). Several skeletal vibrational modes of the first electronically excited state of the neutral species and those of the cation are assigned, with the aid of *ab initio* and density functional calculations. Although ZEKE is governed by the Franck–Condon principle different from direct IR absorption or emission, vibronic coupling in the long ribbon like molecule results in the observation of a few IR active modes. Within the experimental resolution of ~7 cm$^{-1}$, the frequency values from our calculation agree with the experiment for the cation, but differ for the electronically excited intermediate state. Consequently, modeling of the intensity distribution is difficult and may require explicit inclusion of vibronic interactions.

SUBJECT KEY WORDS: line: identification, molecular data; techniques: spectroscopic

## 1. INTRODUCTION

Polycyclic aromatic hydrocarbons (PAH) reveal their presence in the interstellar medium (ISM) by a distinctive set of emission features at 3.3, 6.2, 7.7, 8.6, and 11.3μm (which are also collectively known as the "Unidentified Infrared" [UIR] emission bands).[1] Since their first detection in the planetary nebulae NGC 7027 and BD+30º3639 (Gillett et al. 1973), PAHs have been observed in a wide range of Galactic and extragalactic regions (see Li 2009). Neutral and ionic PAHs have long been suggested to be possible carriers of the 2175Å interstellar extinction bump (Joblin et al. 1992; Li & Draine 2001; Steglich et al. 2010) and the mysterious diffuse interstellar bands (DIBs) (Salama et al. 1999). PAHs are the dominant energy source for the interstellar gas as they provide photoelectrons which heat the gas (Weingartner & Draine 2002). They are also potentially related to the origin of life via the formation of primitive organic molecules including amino acids in the pre-DNA world (Mulas et al. 2006; Bernstein et al. 2002; Shock & Schulte 1990; Wakeham et al. 1980).

Identification of the exact molecular formulas of astronomical PAHs unfortunately has been ineffective so far,[2] largely because of the spectral range issue: the 3.3--17.4μm mid-infrared (MIR) bands are mostly representative of functional groups, not the molecular frame.[3] In contrast,

---

[1] These "UIR" emission features are now generally identified as vibrational modes of PAHs (Léger & Puget 1984; Allamandola et al. 1985): C--H stretching mode (3.3 μm), C--C stretching modes (6.2, 7.7 μm), C--H in-plane bending mode (8.6 μm), and C--H out-of-plane bending mode (11.3 μm). Other C--H out-of-plane bending modes at 11.9, 12.7 and 13.6 μm have also been detected. The wavelengths of the C--H out-of-plane bending modes depend on the number of neighboring H atoms: 11.3 μm for solo-CH (no adjacent H atom), 11.9 μm for duet-CH (2 adjacent H atoms), 12.7 μm for trio-CH (3 adjacent H atoms), and 13.6 μm for quartet-CH (4 adjacent H atoms). Other prominent features are the C-C-C bending modes at 16.4 μm (Moutou et al. 1996) and 17.4 μm (Beintema et al. 1996; Smith et al. 2004, 2007).

[2] Several small PAHs [naphtalene ($C_{10}H_8$), phenanthrene ($C_{14}H_{10}$), pyrene ($C_{20}H_{12}$), perylene ($C_{16}H_{10}$)] have been found in the *Stardust* samples collected from comet Wild 2 (Sandford et al. 2006), and in interplanetary dust particles possibly of cometary origin (Clemett et al. 1993). Although Spencer & Zare (2007) argued that some of the low-mass PAHs seen diffusely on the surfaces along the impact track may be due to impact conversion of aerogel carbon, the majority of the PAHs in the Stardust samples should be of cometary origin (Sandford & Brownlee 2007).

[3] In astronomical modeling, astronomers usually take an empirical approach by constructing "astro-PAH" absorption properties that are consistent with spectroscopic observations of PAH emission in various astrophysical environments (e.g. see Désert et al. 1990, Schutte et al. 1993, Li & Draine 2001, Draine & Li 2001, 2007). The resulting "astro-PAH" absorption cross sections, although generally consistent with laboratory data (e.g. see Fig. 2 of Draine & Li 2007), do not represent any specific material, but approximate the actual absorption properties of the PAH mixture in astrophysical regions.

the far infrared (FIR) bands are considered sensitive to the skeletal characteristics of a molecule. Hence they contain specific fingerprint information on the identity of a PAH molecule and could lead to the identification of individual PAHs (Langhoff 1996; Zhang et al. 1996; Joblin et al. 2009)

One mission of the *Stratospheric Observatory for Infrared Astronomy* (SOFIA) and the *Herschel Space Observatory* is thus to target the FIR and sub-millimeter (submm) wavelength region of the ISM. The goal is to answer the questions: "*What is the molecular makeup of the ISM and how does that relate to the origin of life?*" To assist with data interpretations of these missions, laboratory experimental data of PAHs in the same spectral region are in urgent need.

Experiments in the FIR are typically challenging due to the lack of adequate light sources and sensitive detectors. Moreover, spectroscopy of ions and more so of ions of non-volatile species, is further hindered by the low achievable particle density. We have recently succeeded in combining the laser desorption (LD) technique with the zero kinetic energy photoelectron (ZEKE) spectroscopy technique (see §2 for a brief description; for more details see Zhang et al. 2008). With this combination, albeit still challenging, we offer an alternative approach in solving the issues of FIR spectroscopy of non-volatile cations.

There have been several reports of the ZEKE spectroscopy of small PAHs including naphthalene ($C_{10}H_8$), anthracene ($C_{14}H_{10}$), and tetracene ($C_{18}H_{12}$) (Cockett & Kimura 1994; Cockett et al. 1993; Zhang et al. 2008). In this paper, we report in §4 the ZEKE spectroscopy in the FIR of pentacene ($C_{22}H_{14}$), a five-ring PAH molecule. Detailed spectroscopic analysis for the vibrational levels of the first electronically excited state ($S_1$) and the ground cationic state ($D_0$) will be discussed, with the assistance of *ab initio* and density functional calculations (§5). Structural changes due to electronic excitation and ionization will be elucidated from the observed active vibrational modes. In particular, we report several IR active modes that are important for astronomy (§5.4). Comparisons between the observed modes of pentacene and smaller catacondensed PAHs, particularly tetracene, will be discussed. Readers who are interested only in the resulting FIR spectra may wish to proceed directly to §5.4.

## 2. The ZEKE Technique

The ZEKE technique has been introduced to the gas phase spectroscopy community for more than two decades (Schlag 1998). Typically in a ZEKE experiment, a tunable laser (or lasers in a multiphoton excitation scheme) excites valence electrons into high Rydberg states just below the ionization threshold of a selected vibronic state of the cation. These Rydberg state electrons can orbit around a cationic core for a long time (micro to milliseconds), allowing prompt electrons directly ionized from the laser to escape the detection region. With a delayed electric pulse, the Rydberg state species can then be field ionized to generate ZEKE electrons. Thus only when the excitation energy is close to a particular vibronic state of the cation can Rydberg states exist and ZEKE electrons be detected.

The spectral resolution of ZEKE is determined by the delayed pulsed electric field and the excitation laser, not affected by the discrimination ability of threshold electrons as in typical threshold photoelectron spectroscopy experiments (Ng 1991). Usually in the wavenumber range, resolutions of ZEKE experiments on the order of ~100 kHz ($3\times10^{-6}$ cm$^{-1}$) are achievable with careful controls of experimental conditions and narrow line-width lasers (Merkt & Schmutz 1998; Osterwalder & Merkt 1999; Schlag 1998).

The high Rydberg states have the same vibronic properties as those of the corresponding cation, therefore ZEKE is ideal for the study of vibrational spectroscopy of cations. In addition, Rydberg states are longer lived when they are associated with lower vibronic states of the cation, which makes ZEKE particularly suitable for studies of lower frequency vibrational modes. Thus by detecting pulsed field ionized electrons in ZEKE spectroscopy, we can avoid the detector problem in FIR and submm spectroscopy.

There have been several reports of ZEKE spectroscopy of small PAHs including naphthalene ($C_{10}H_8$), anthracene ($C_{14}H_{10}$), and tetracene ($C_{18}H_{12}$) (Cockett & Kimura 1994; Cockett et al. 1993; Zhang et al. 2008). The difficulty in vaporizing large PAHs has severely limited further efforts. Fortunately, laser desorption has recently resolved the issue and offered internal cooling of the neutral species (Mons et al. 2002; Nir et al. 2000).

In our LD setup, a low power pulsed IR laser is used for vaporization of the solid sample. Due to the fast efficient heating of the pulsed laser, only neutral intact molecules are ejected into the gas phase. The neutral molecules are entrained in a supersonic jet of argon for rotational and vibrational cooling. The cold isolated neutral species enters into the detection region where they can be first pumped to an electronically excited state by an ultraviolet (UV) laser then ionized by a scanning second laser. This *resonantly enhanced multiphoton ionization* (REMPI) scheme ultimately generates cations in different vibrational levels. Thus by using tunable UV lasers in ZEKE via resonant excitation and ionization, we bypass the light source problem in typical FIR and submm experiments. The combination of ZEKE and LD has been demonstrated in our studies of tetracene (Zhang et al. 2008).

## 3. Experiment Setup

The experimental apparatus has been described in detail in a previous publication (Zhang et al. 2008). Briefly, it consists of a differentially pumped high vacuum system with a laser desorption source and a time of flight mass spectrometer (TOF-MS), which could be converted into a pulsed field ionization (PFI) zero kinetic energy photoelectron spectrometer. The pentacene sample (Aldrich) was smudged onto a graphite rod by hand. The desorption laser (Spectra Physics GCR 230) at 1064 nm with a pulse energy of ~0.1 mJ/pulse was focused onto the rod by a lens with a focal length of 6". The desorbed pentacene vapor was entrained in a supersonic expansion of argon with a stagnation pressure of 3 atm and a pulsed nozzle of 1 mm in diameter. The laser systems for the REMPI experiment included a Nd:YAG (Precision II 8000, Continuum) pumped optical parametric oscillator (OPO, Panther, Continuum) and a Nd:YAG (Spectra Physics GCR 190) pumped dye laser (Laser Analytical Systems. LDL 20505). The OPO laser in the 523 - 538 nm range had a pulse energy of ~1.5 mJ/pulse. The ionization laser in the 280 - 290 nm range, obtained from the frequency doubled dye laser system, had a pulse energy of ~0.8 mJ/pulse. The absolute wavelength of each laser was calibrated using an iron hollow-cathode lamp filled with neon. The pump laser and ionization laser were set to counter propagate, and the light path, the flight tube, and the molecular beam were mutually perpendicular. The relative timing among the three laser pulses was controlled by two delay generators (Stanford Research, DG535), and the optimal signal was obtained under temporal overlap between the pump and ionization lasers. In the ZEKE experiment, molecules excited to high Rydberg states stayed in the excitation region for 1 - 2 μs in the presence of a constant DC spoiling field of ~ 1 V/cm. Further ionization and extraction was achieved by a pulsed electric field of ~8 V/cm.  The DC spoiling field in the ionization region was to remove prompt electrons generated from direct photoionization.

Gaussian 03 suite was used to optimize the molecular structure and to obtain vibrational frequencies for assignment of the observed vibronic structures from REMPI and ZEKE (Frisch et al. 2003).  For the ground state of the neutral ($S_0$) and the cationic state $D_0$, density functional theory (DFT) calculations using the B3LYP functional were performed with the 6-31G+ (dp) basis set.  The excited state $S_1$ was calculated at the CIS level using the 6-31G+ (dp) basis set. Due to the fact that vibrational frequencies generated by *ab initio* calculations are usually too high, a general practice is to use a scaling factor to match the experimental results (Casida et al. 1998).  For the $S_1$ state of pentacene, we used a scaling factor of 0.92, but no scaling factor was necessary for the $D_0$ state from our DFT calculation.

## 4. Results

### 4.1. Two-color 1+1' REMPI spectrum

The two-color 1+1' REMPI spectrum of pentacene near the origin of the $S_1 \leftarrow S_0$ electronic transition is displayed in Figure 1.  The ionization laser was set at 280 nm and was temporally overlapped with the scanning resonant laser.  Since one photon from each laser was used, and the overall process involved the concurrent excitation of both lasers, this multiphoton process is typically designated as 1+1' REMPI in laser spectroscopy.  The intense peak at 18657 cm$^{-1}$ is assigned as the origin band, and the other observed vibronic transitions are listed in Table 1.  The labeling of each vibrational mode is based on spectroscopic conventions, i. e. by using consecutive numbers in reference to the symmetry species and the frequency in decreasing order. The calculated frequencies for the $S_1$ state, after scaling by a factor of 0.92, agree with the observed results within 22 cm$^{-1}$.

Vibronic activities in REMPI are controlled by the Franck-Condon principle and affected by vibronic coupling.  For linear polyacenes, all vibrational levels of $A_g$ modes and levels with even quantum numbers of other modes are symmetry allowed, and odd levels of $B_{3u}$ and $B_{1g}$ modes are only allowed via vibronic coupling with higher excited states.  The intensity of the symmetry allowed vibronic bands is determined by the overlap in the vibrational wavefunctions of the related electronic states.  In Table 1, with the exception of mode 18, the rest of the observed bands are due to vibronic coupling.  The intensities of these bands are comparable to or even stronger than the Franck-Condon allowed band.

The displacement vectors of the observed modes are shown in Figure 2.  The assignment is based on symmetry analysis as well as calculation results.  All current assignments agree with those proposed by Griffiths *et al* (Griffiths & Freedman 1982).  Our own theoretical result contains only one $A_g$ mode below 500 cm$^{-1}$ after scaling.  Hence the assignment of mode 18 corresponding to in-plane longitudinal stretching near 260 cm$^{-1}$ is unambiguous.  Interestingly the same mode has also been observed in tetracene, anthracene, naphthalene and benzene (Garforth et al. 1948; Beck et al. 1980; Amirav et al. 1981; Lambert et al. 1984).

The assignment of the other three bands in Figure 1 is guided by the vibronic coupling principle.  All three bands are fundamental transitions of Franck-Condon forbidden out-of-plane modes.  For mode 33, although the experimental result differs from calculation by 10 cm$^{-1}$, the assignment is confirmed based on two facts.  First, it is the only possibility within 25 cm$^{-1}$ according to the theoretical calculation result; and second, this assignment can be confirmed from the ZEKE spectrum, as will be seen in the following section.  The third peak in the REMPI spectrum at 199 cm$^{-1}$ is 33 cm$^{-1}$ higher than the 0-2 transition of mode 33, and this difference is too large and is opposite to the trend of typical anharmonicity.  Furthermore, there is no other symmetry allowed mode in this region from our calculation.  Thus this peak at 199 cm$^{-1}$ has to be the fundamental transition of mode 101 with $B_{3u}$ symmetry.  The last peak at 342 cm$^{-1}$ is assigned as mode 100 with $B_{3u}$ symmetry, consistent with the trend of the observed vibrational modes.

We attribute the violation of the Franck-Condon selection rule in pentacene to the decreased rigidity of the molecular structure.  In polyacenes, the rigidity of the ribbon decreases with the increasing number of fused rings.  Deformation of the molecular frame lowers the symmetry group of the molecule, relaxing the selection rule and activating forbidden modes.  While tetracene was observed to strictly obey the selection rule (Zhang et al. 2008), pentacene is quite different.  If we define the length of the ribbon as *L*, in Figure 2 among the observed waving modes, the waving cycles are changed from *2L/3* in mode 33 to *L/2* in mode 101 and to *L/3* in mode 100. If we further follow the *2L/n* pattern, where *n* is an integer, the mode with *n = 5* would correspond to $B_{1g}$ symmetry at 282 cm$^{-1}$. It is missing in the observed spectrum, probably due to the limited signal to noise ratio.

### 4.2. ZEKE spectra

By scanning the ionization laser while setting the resonant laser at one of the intermediate states identified in the above REMPI experiment, we obtained pulsed field ionization ZEKE spectra as shown in Figure 3. The assignment of the vibrational levels of the cation is noted by a superscript "+". The experimental and theoretical values are shown in Table 2. The calculation was performed at the B3LYP/6-31G+ (dp) level with no scaling factor for the vibrational frequencies of the cation. Limited by the linewidth of the resonant transitions and the pulsed electric field, the uncertainty of the experimental values of the ZEKE spectra is 7 cm$^{-1}$. From trace 3a recorded via the origin of the $S_1$ state, the adiabatic ionization potential is determined to be $53266 \pm 7$ cm$^{-1}$, taking into account the shift induced by the delayed electric field. This value agrees with the result of 6.589 ±0.001 eV by Gruhn *et al* (Gruhn et al. 2002).

Overall the spectra in Figure 3 are sparse, but they are not dominated by one vibrational band. Mode 102 is not observed in the REMPI spectrum and new to ZEKE. Its displacement vectors are shown in Figure 2. This mode is the lowest in frequency among the $B_{3u}$ modes of the cation, corresponding to a complete waving cycle (*n = 2*) in the out of plane vibrational motion. Mode $101^+$ at 187 cm$^{-1}$ and mode $102^+$ at 38cm$^{-1}$ are only observable as a combination band $101^+102^+$ at 225 cm$^{-1}$, and its presence is almost ubiquitous, observable in all three spectra (a – c). We did not observe any appreciable vibrational bands via the intermediate level $100^1$, probably due to its low intensity in the REMPI spectrum (Fig. 1).

The agreement between theory and experiment in the ZEKE spectra can help to confirm the assignment of the intermediate vibrational levels of the $S_1$ state. In our previous studies of substituted aromatic compounds (He et al. 2004a, b, c; Wu et al. 2004), we have observed a propensity rule where the vibrational excitation of the intermediate state is preserved during ionization. Although this propensity rule is not prominent in this case, some correlation is still expected. For example, the assignment of mode 33 in the ZEKE spectrum is definitive, while the assignment of the corresponding band in the REMPI spectrum has a difference of 10 cm$^{-1}$ between theory and experiment (Table 1). The ambiguity in the assignment of the REMPI spectrum can be removed when we consider the propensity of maintaining the same mode and level of vibrational excitation during ionization.

## 5. Discussion

### 5.1. Vibrational modes of the $S_1$ state

The observed mode structure should be reflective of the geometry changes upon electronic excitation. Table 3 lists the calculated geometric parameters for the three related electronic states of pentacene. The numbering scheme of the carbon atoms is shown in the top panel of Figure 4. The most significant changes upon electronic excitation and ionization are the length and width of the ribbon; both dimensions contract upon excitation and extend back close to their original values upon ionization. These changes are in accord with the activation of the longitudinal stretching mode 18 in the spectrum. Interestingly, among the polyacenes we have investigated, this sequence of geometry change seems ubiquitous, and mode 18 has a universal presence in the REMPI spectrum of these polyacenes.

Some of the changes in bond length listed in Table 3 can be qualitatively explained by a simple Hückel calculation. Figure 4 shows the lowest unoccupied molecular orbital (LUMO) and the highest occupied molecular orbital (HOMO) from our simple Hückel calculation. The most dramatic change between the two orbitals is the disappearance of the horizontal nodal plane in the LUMO. Thus in the $S_0 \rightarrow S_1$ transition when an electron is excited from the HOMO to the LUMO, the width of the ribbon should shrink due to the increasing bonding nature along the short axis. This result is in agreement with the higher level calculation listed in Table 3 where the bond lengths of C1 – C22, C3 – C20 and C5 – C18 all decrease from $S_0$ to $S_1$. The shift of nodal planes in the longitudinal direction also incurs further changes. The disappearance of the nodal planes between C2 and C3 as well as C4 and C5 decreases the bond lengths of C2 – C3 and C4 – C5.

The rigidity of the molecule is expected to decrease with the elongation of the ribbon. In fact, the frequency of the $A_g$ longitudinal stretching mode shifts to lower values with the increase of the number of aromatic rings: from 923 cm$^{-1}$ for benzene (Garforth et al. 1948), 501 cm$^{-1}$ for naphthalene (Beck et al. 1980), 390 cm$^{-1}$ for anthracene (Lambert et al. 1984), 307 cm$^{-1}$ for tetracene (Amirav et al. 1981), to 263 cm$^{-1}$ for pentacene. Moreover, the selection rule becomes relaxed, and the Franck-Condon forbidden but vibronic allowed bands show increasing intensities with the elongation of the molecular frame. The $S_1$ state of anthracene exhibits weak but yet observable activities of several $B_{1g}$ modes. For tetracene, additional weak $A_u$ modes are also recognizable. In pentacene, the Franck-Condon forbidden modes are nearly equal in intensity or even stronger than that of the $A_g$ mode.

The flexibility of the pentacene ribbon is a subtle effect and challenges our theoretical calculation methods. Our CIS results indicate that the first excited state of pentacene is slightly nonplanar, and the C-C bond in the central ring is likely weakened by this geometric deformation. We then conducted the Franck-Condon calculation associated with $D_0 \leftarrow S_1$ and $S_1 \leftarrow S_0$ transitions. However, the resulting vibronic structures are still dominated by $A_g$ modes; the intensities of which are several orders higher than those of the Franck-Condon forbidden modes. This disagreement illustrates the inaccuracy in the geometry of the first excited state. Moreover, simple Hückel calculation resulted in four vertical nodal planes in the HOMO and six vertical nodal planes in the LUMO. Thus we would intuitively expect that during the $S_1 \leftarrow S_0$ transition, the ribbon should elongate along the long axis, which is opposite from our higher level CIS calculation (Table 3). We suspect that explicit inclusion of vibronic coupling is needed to even qualitatively reproduce the experimental intensity and hence the molecular geometry.

### 5.2. Vibrational modes of the cation

Although the sparse ZEKE spectra of all reported polyacenes are indicative of small changes in geometry upon ionization from the electronically excited intermediate state, the spectra in Fig. 3 do not show clear propensity in preserving the vibrational motion of the intermediate $S_1$ state. In fact, the geometry of the cation is more similar to that of the neutral ground state than that of the first excited state. In the ionization step based on our calculation, the molecular frame changes from slightly non-planar to planar, activating the $B_{3u}$ mode. From a simple intuitive consideration, ionization involves removal of the electron from the LUMO, which leaves a single electron in the HOMO. Some degree of recovery in the bonding nature of the $S_0$ state is therefore expected.

The frequency of the $A_g$ longitudinal stretching mode of the cation shares the same trend as that of the $S_1$ state, with decreasing values from anthracene to pentacene. The flexibility of the ribbon is also the most prominently manifested in the ZEKE spectra of pentacene. For anthracene and tetracene, the modes observed in ZEKE are identical to those observed in REMPI. For pentacene, however, a new mode $102^+$ is observed in ZEKE from the combination band $101^+102^+$. Calculation of the ZEKE spectra based on the geometry of the $S_1$ state, on the other hand, cannot reproduce the relative intensities of the observed vibrational modes, due to the inaccuracy of the theoretical structure of the $S_1$ state.

### 5.3. Remarks on the comparison between experimental and theoretical vibrational frequencies

The quality of agreement between experiment and theory from Tables 1 and 2 is quite different. In Table 2, the experimental vibrational frequencies of the cation agree with the unscaled calculation results within the experimental error, while in Table 1, after a scaling factor of 0.92, the disagreement is still obvious. In addition, there is no reliable trend in the disagreement; while the theoretical value is larger than the experimental value at the two extremes of the spectrum (mode 33 and 100), the opposite is true for the middle section of the spectrum (mode 101 and 18). This result highlights the limit of the prediction power of calculation: while we can rely on the DFT results for the ground state, the calculation of excited electronic states is still problematic. This situation is in accord with the analysis of §5.1, where violation of the Franck-Condon selection rule and inconsistency in the geometric changes upon electronic excitation and ionization were noted.

The reliability of the frequency calculation for the ground state of the cation can be used to predict the frequency of other modes unobservable from ZEKE. Cationic pentacene has 15 IR active modes with non-negligible intensities below 1000 cm$^{-1}$, as listed in Table 4. Hudgins & Allamandola reported the IR spectroscopy of pentacene cation using the matrix isolation method covering the near- and mid-IR region (740 cm$^{-1}$ to 1550 cm$^{-1}$) (Hudgins & Allamandola 1995). The authors reported four modes at 741, 749, 862, and 934 cm$^{-1}$, none of which were observable in our experiment but were obtained from our calculation in Table 4. The experimental frequencies from the work of Hudgins & Allamandola agree with our calculation within 23 cm$^{-1}$. Together with our results, the two complementary experiments span 1500 cm$^{-1}$ containing five observed experimental values (bold-phased in Table 4).

The structural information from calculation should be treated with caution. The REMPI and ZEKE spectra are dominated by vibronic coupling, and our CIS calculation has failed to capture the accurate structure of the molecular frame. Calculations of IR transition intensities and rotational constants therefore await further improvements in theoretical modeling, most likely with the explicit inclusion of vibronic coupling (Burrill et al. 2004; Goode et al. 1997; Johnson 2002a,b).

### 5.4  Astrophysical implications

In Figure 5 we display the calculated FIR spectra of cationic naphthalene, anthracene, tetracene, and pentacene with DFT using the 6-31G+(dp) basic set. It is apparent that their FIR spectroscopy, particularly in the range with λ > 30 μm, differs considerably from one PAH species to another. In contrast, their mid-IR modes do not distinguish themselves as much (e.g. see Allamandola & Hudgins 2003). This is because the mid-IR bands are mostly representative of functional groups, not the molecular frame, while the FIR bands arise from bending of the whole PAH skeleton (mostly out-of-plane) and are thus intrinsically related to the molecular structure (e.g. see Berné et al. 2009, Mattioda et al. 2009, Joblin et al. 2009, Ricca et al. 2010).

One can also see from Figure 5 that the FIR bands are much weaker than the mid-IR vibrational bands and thus making them more difficult to detect in space. This is unfortunate since PAHs tend to release their vibrational energy mostly through mid-IR emission (e.g. see Li 2004, Berné et al. 2009, Mattioda et al. 2009, Joblin et al. 2009, Ricca et al. 2010). However, with the advent of space-borne and airborne FIR telescopes and the ground-based *Atacama Large Millimeter Array* (ALMA), particularly in view of the prospects of much improved sensitivity with future space instruments, it is not impossible that individual PAH molecules may be identified in space (e.g. see Mulas et al. 2006). We also plot in Figure 5 the wavelength coverages of some key space telescopes as well as ALMA which may allow one to explore the identification of specific PAH molecules in space through their FIR bands.  Although only two IR active modes of cationic pentacene at 53.3 and 266 μm have been observed in this work, all three IR active modes including the one at 84.8 μm may be detectable by space missions such as the SAFARI instrument on board SPICA.

Finally, we should note that the four PAH species (naphthalene, anthracene, tetracene, and pentacene) shown in Figure 5 are all ribbon-like catacondensed molecules. In the Galactic and extragalactic ISM, the typical PAHs may be larger and have a compact, pericondensed structure (see Figure 1 of Li 2009), as small PAHs with an open structure may not survive under the harsh interstellar conditions (see Tielens 2008).[4] Nonetheless, a careful compilation of the FIR vibrational frequencies of small PAHs would at least provide a very useful complement to the NASA/Ames mid-IR spectroscopy database of PAHs of a range of sizes (Bauschlicher et al. 2010).  In addition, several pericondensed PAHs are under investigation in our laboratory using the same technique.

## 6.  Conclusion

Spectroscopic properties of the electronically excited and ionic states of pentacene have been studied using REMPI and ZEKE. We have identified a total symmetric $A_g$ mode corresponding to stretching of the long axis. The decreasing vibrational frequency of this mode from naphthalene to pentacene signifies the increasing flexibility of the molecular frame. For the cation, we have also observed two IR active modes in a combination band, which related to the out-of-plane waving motion of the molecular frame and may be detectable by space missions such as the SAFARI instrument on board SPICA. Simple Hückle calculations can partially explain the changes in geometry regarding the addition or elimination of nodal planes in the LUMO and HOMO. However, it assumes planar geometry and hence fails to include any effect of out-of-plane deformation. From comparisons between theoretical and experimental results, we have concluded that frequency values from DFT using the B3LYP functional with the 6-31G+ (dp) basis set for the ground state of the cation are reasonably reliable, hence frequency information relevant for astronomy can be retrieved with some confidence. However, vibronic coupling dominates the spectra, and the reliability of theoretical intensity and geometric parameters awaits further theoretical modeling.

---

[4]  However, naphthalene has been detected in the Stardust cometary sample (Sandford et al. 2006), although some uncertainties still remain (Spencer & Zare 2007 and Sandford & Brownlee 2007).

ZEKE spectroscopy is particularly sensitive to low frequency modes of the molecular frame, which gives it a superior advantage in fingerprinting the skeletal motions of the molecular species. The combination of LD and ZEKE has the potential of being an important tool in astrophysics. Although ZEKE follows a different selection rule from IR spectroscopy, and in the current experiment, it only samples a limited number of vibrational modes, it can bench mark a few of the observed modes, guiding theoretical developments and refining theoretical predictions.

7. Acknowledgement

We are grateful to our talented machinist, Ted Hinke, for his designs and modifications of the experimental apparatus. This work is supported by the National Aeronautics and Space Administration under award No. NNX09AC03G. Aigen Li is supported in part by the NSF grant AST 07-07866, a Spitzer Theory grant and a Herschel Theory grant.

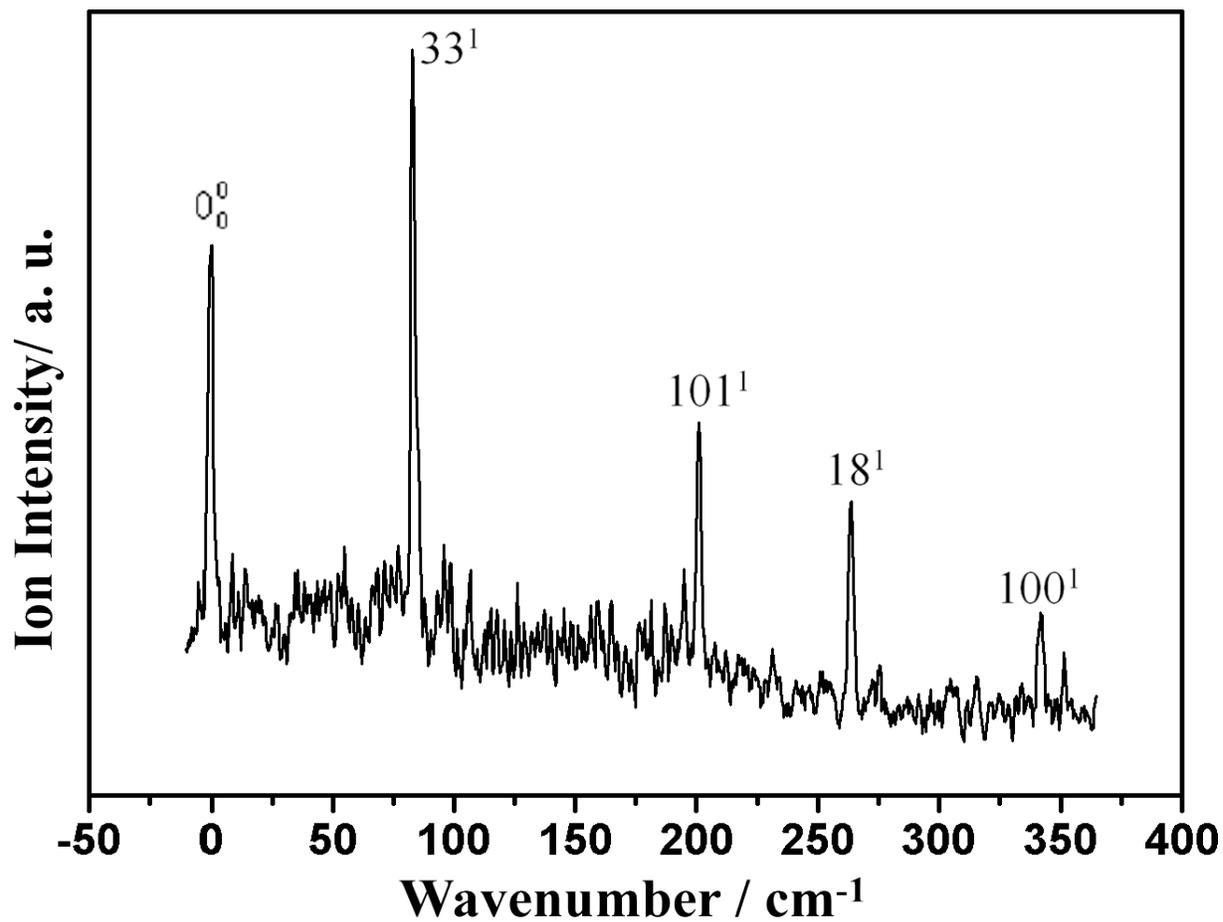

**Figure 1** 1+1' REMPI spectrum of pentacene. The spectrum is shifted by 18657 cm$^{-1}$ (the origin of the $S_1 \leftarrow S_0$ transition) to emphasize the frequencies of the different vibrational modes of the $S_1$ state.

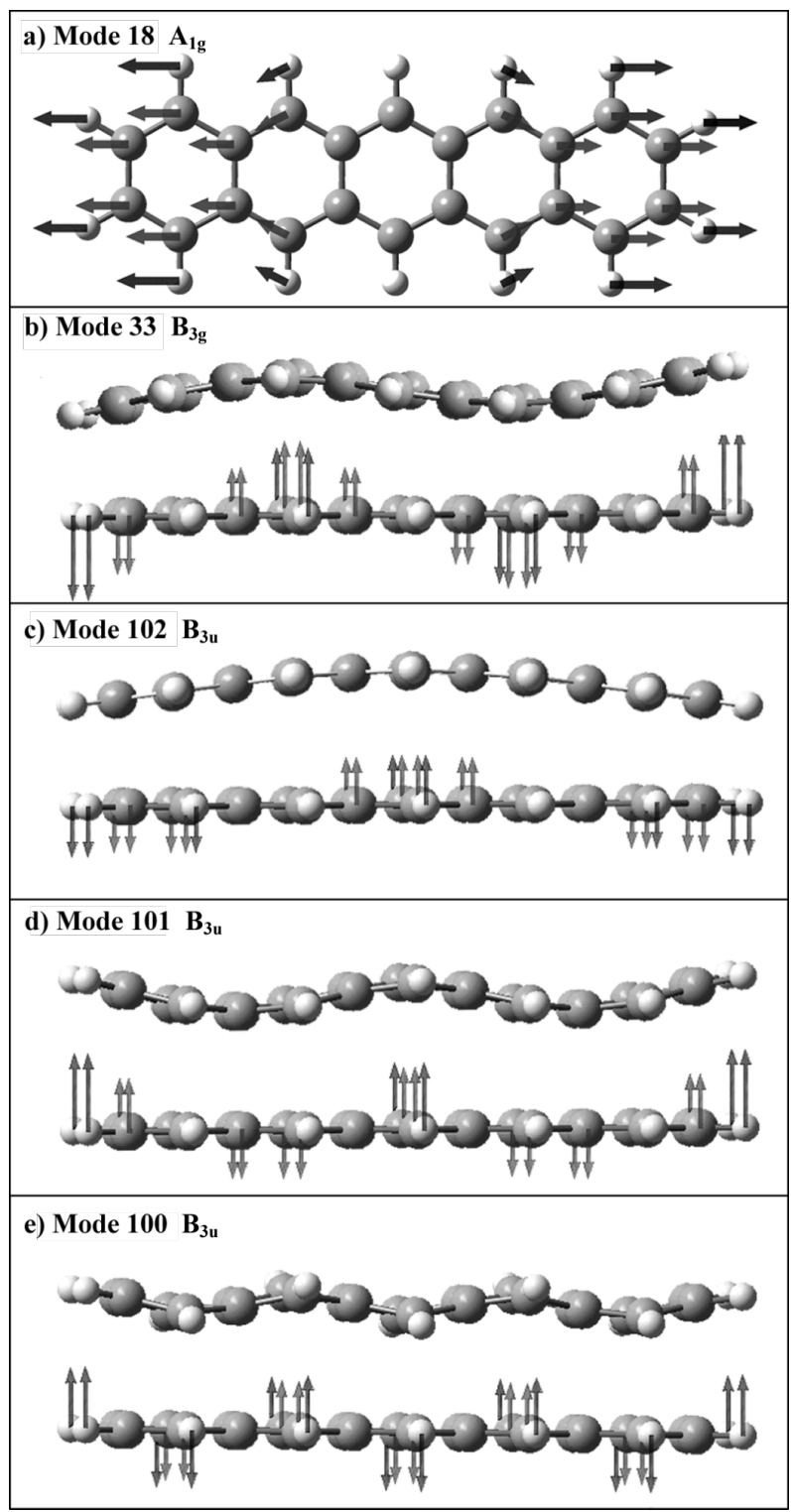

a) Mode 18 $A_{1g}$

b) Mode 33 $B_{3g}$

c) Mode 102 $B_{3u}$

d) Mode 101 $B_{3u}$

e) Mode 100 $B_{3u}$

**Figure 2** Displacement vectors for five normal modes of pentacene. Mode 102 is not observed in the REMPI spectrum of Fig. 1 but observed in the ZEKE spectrum of Fig. 3.

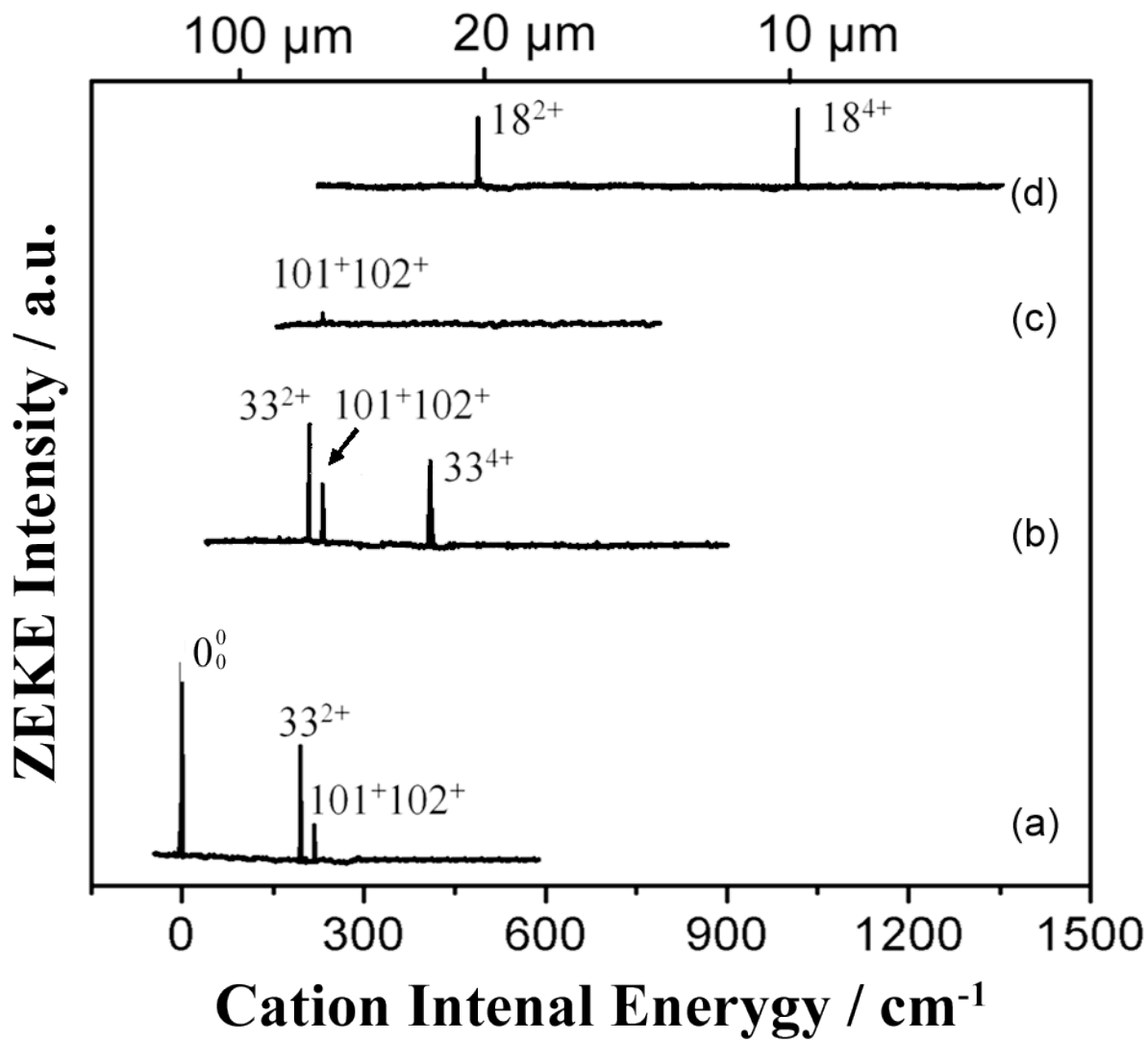

**Figure 3** Two-color ZEKE spectra of pentacene recorded via the following vibrational levels of the $S_1$ state as intermediate states: (a) $0_0^0$, (b) $33^1$, (c) $101^1$, (d) $18^1$. The energy in the abscissa is relative to the ionization threshold at 53266 $cm^{-1}$. The assignment in the figure refers to the vibrational levels of the cation.

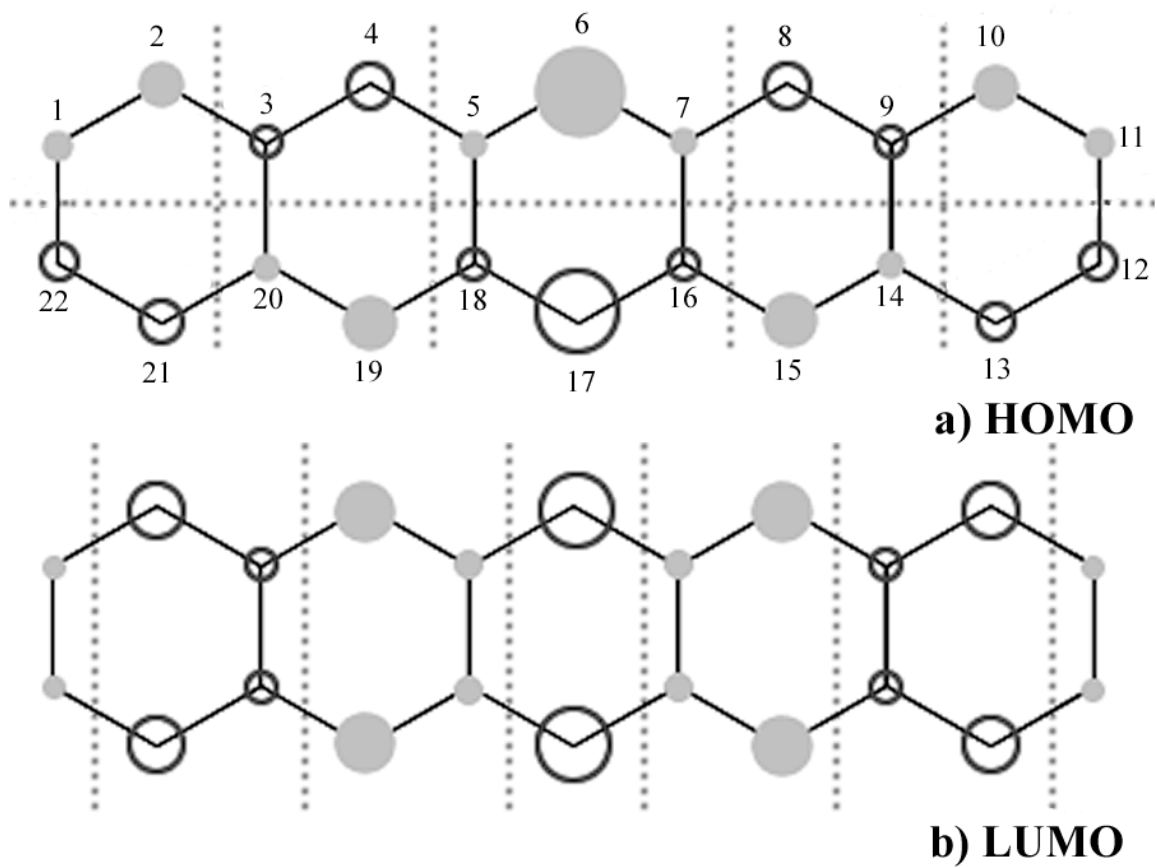

**Figure 4** The (a) HOMO and (b) LUMO of pentacene with the numbering scheme of the carbon atoms. The dashed lines mark the nodal planes based on a simple Hückel calculation.

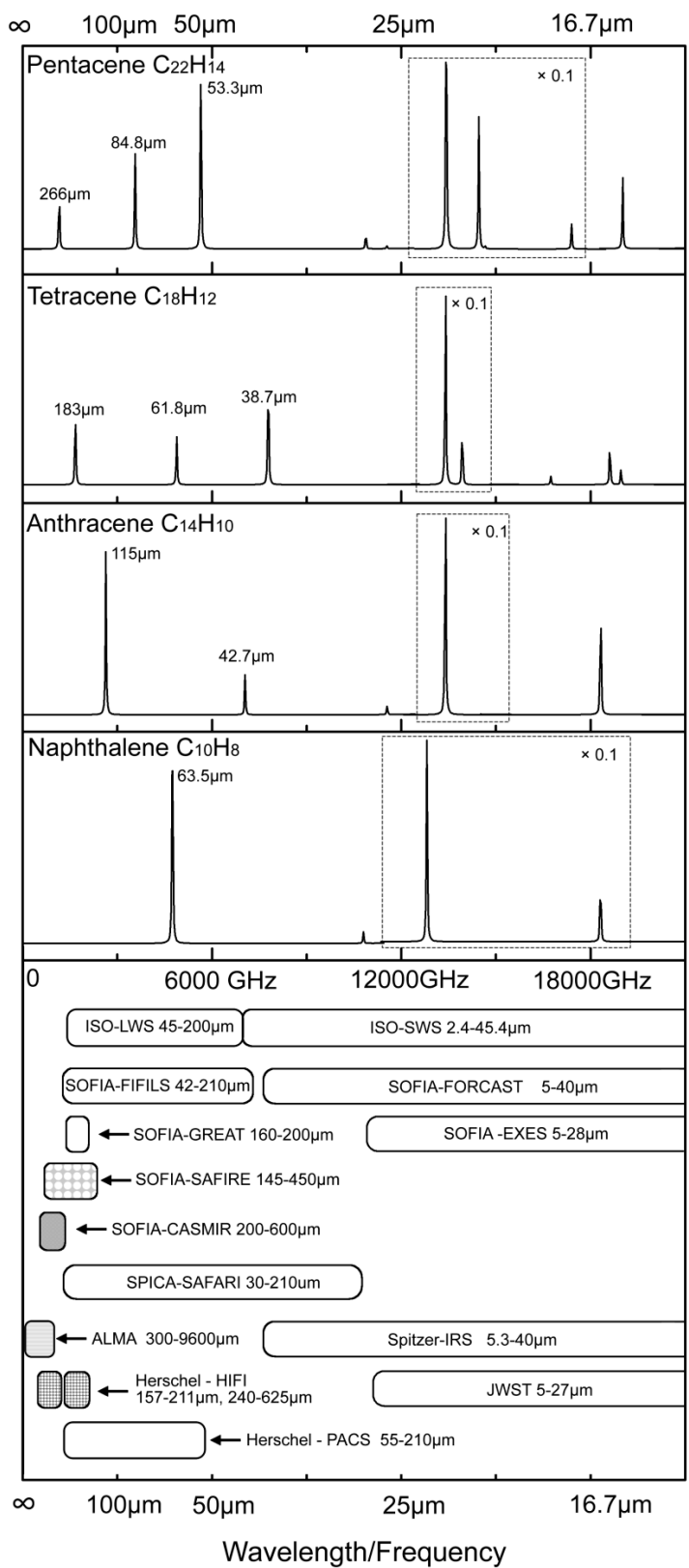

**Figure 5** The calculated IR spectroscopy of naphthalene, anthracene, tetracene, and pentacene. Also shown are the wavelength coverages of some key astronomical instruments which are relevant for detecting individual PAHs in space: (1) the *Heterodyne Instrument for the Far-Infrared* (HIFI; 157--211μm, 240--625 μm) and the *Photodetector Array Camera and Spectrometer* (PACS; 55--210 μm, 240--625 μm ) on board *Herschel*, (2) the *SPICA Far-Infrared Instrument* (SAFARI; 30--210 μm ), a FIR imaging spectrometer on board the *Space Infrared Telescope for Cosmology and Astrophysics* (SPICA); (3) *ALMA* (0.3--9.6 mm); (4) the *Long Wavelength Spectrometer* (LWS; 45--200 μm ) and the *Shorter Wavelength Spectrometer* (SWS; 2.4—45.4 μm) on board the *Infrared Space Observatories* (ISO); and (5) the *Caltech Submillimeter Interstellar Medium Investigations Receiver* (CASMIR; 200--600 μm ), the *Field Imaging Far-Infrared Line Spectrometer* (FIFILS; 42--210 μm ), the *German Receiver for Astronomy at Terahertz Frequencies* (GREAT; 160--200 μm ), and the *Submillimeter and Far-InfraRed Experiment* (SAFIRE; 145--450 μm ) on board *SOFIA*. Also shown are the wavelength coverages of the *Infrared Spectrograph* (IRS; 5.3--40 μm ) on board the *Spitzer Space Telescope* and the *Mid-Infrared Instrument* (MIRI; 5--27 μm ) on board the *James Webb Space Telescope* (JWST), the *Faint Object Infrared Camera for the SOFIA Telescope* (FORCAST; 5--40 μm ) and the *Echelon-Cross-Echelle Spectrograph* (EXES; 5--28 μm ) on board SOFIA .

Table 1. Observed and calculated vibrational frequencies of the $S_1$ state of pentacene[*]

| Exp | Calc | Assignment (Mode, symmetry) |
|-----|------|------------------------------|
| 83 | 93 | $33^1$, $B_{1g}$ |
| 199 | 177 | $101^1$, $B_{3u}$ |
| 263 | 256 | $18^1$, $A_g$ |
| 342 | 352 | $100^1$, $B_{3u}$ |

[*]A scaling factor of 0.92 is included in the calculation result

Table 2. Observed and calculated vibrational frequencies of pentacene cation

| $0_0^0$ | $33_0^1$ | $101_0^1$ | $18_0^1$ | Cal | assignment |
|---|---|---|---|---|---|
| 0 | | | | 0 | $0^{0+}$ |
| 196 | 198 | | | 198 | $33^{2+}$ |
| 219 | 221 | 221 | | 225 | $101^+102^+$ |
| | 396 | | | 396 | $33^{4+}$ |
| | | | 520 | 522 | $18^{2+}$ |
| | | | 1047 | 1044 | $18^{4+}$ |

Table 3. Molecular geometry parameters of pentacene in the $S_0$, $S_1$, and $D_0$ states

| Bond length (Å) | $S_0$ | $S_1$ | $D_0$ |
|---|---|---|---|
| C1 – C2 | 1.367 | 1.369 | 1.378 |
| C2 – C3 | 1.437 | 1.416 | 1.424 |
| C3 – C4 | 1.390 | 1.409 | 1.405 |
| C4 – C5 | 1.416 | 1.389 | 1.406 |
| C5 – C6 | 1.404 | 1.404 | 1.407 |
| C1 – C22 | 1.434 | 1.412 | 1.421 |
| C3 – C20 | 1.456 | 1.427 | 1.448 |
| C5 – C18 | 1.458 | 1.453 | 1.456 |
| Distance (Å) | | | |
| C1 – C11 | 12.247 | 12.210 | 12.236 |
| C2 – C21 (C4 – C19) | 2.823 (2.818) | 2.790 (2.792) | 2.824 (2.821) |
| (C6 – C17) | (2.820) | (2.780) | (2.824) |

Table 4. IR active FIR modes of cationic pentacene from DFT calculation

| Frequency/ cm$^{-1}$ | Intensity | Frequency/ cm$^{-1}$ | Intensity | Frequency/ cm$^{-1}$ | Intensity |
|---|---|---|---|---|---|
| **38** | **0.62** | 481 | 12.93 | **766** | **100.00** |
| 118 | 0.88 | 488 | 0.28 | **879** | **15.99** |
| **187** | **1.89** | 579 | 2.19 | 916 | 1.37 |
| 362 | 0.17 | 634 | 0.63 | **957** | **73.92** |
| 447 | 31.27 | **753** | **2.44** | 994 | 13.92 |